\documentclass[aps,showpacs,floatfix,twocolumn]{revtex4}
\usepackage{amsmath}
\usepackage{epsfig, graphicx}
\begin{document}

\title{Theory of Antineutrino Monitoring of Burning MOX Plutonium Fuels}

\author{A.C. Hayes, H.R. Trellue, Michael Martin Nieto, and W.B. Wilson}
\affiliation{ Los Alamos National Laboratory, Los Alamos, NM  87545}
\date{\today}

\begin{abstract}
%%%%%%%%%%%%%%% LA-UR Rotated Box %%%%%%%%%%%%%%
%\hspace*{-1.3in}
%\rotatebox{90}{%
%\fbox{\parbox[t]{1.0in}{LA-UR-11-11437}}}
%\vspace*{-1.05in}
%%%%%%%%%%%%%%%%%%%%%%%%%%%%%%%%%%%%%%%%

This letter presents the physics and feasibility of reactor antineutrino monitoring to verify the burnup of plutonium loaded in the reactor as a Mixed Oxide (MOX) fuel. It examines the magnitude and temporal variation in the antineutrino signals expected for different MOX fuels, for the purposes of nuclear accountability and safeguards.  The antineutrino signals from reactor-grade and weapons-grade MOX are shown to be distinct from those from burning low enriched uranium. Thus, antineutrino monitoring could be used to verify the destruction of plutonium in reactors, though verifying the grade of the plutonium being burned is found to be more challenging.
\end{abstract}
\maketitle

\section{Introduction}
Mixed oxide (MOX) fuels refer to reactor fuel that contain oxides of more than one fissionable actinide, and typically
involve a mixture of plutonium  and natural or depleted uranium. When the plutonium makes up about 5\% of the total MOX fuel, the fuel burns with a  similar reactivity to that of normal Low Enriched uranium (LEU) fuels. 
One major
motivation in burning MOX fuels is the destruction of either weapons-grade or reactor-grade
plutonium. However, such schemes raise new nuclear safeguards concerns. For example, is it possible to verify how much MOX fuel was loaded, and is it possible
to verify the grade of plutonium being burned?
A number of current and past studies \cite{Borovoi1} have examined antineutrino monitoring of reactors to 
verify the reactor power and  the isotopic content of the fuel.
Here we examine the feasibility of using
antineutrino monitoring to verify the burning of plutonium in MOX fuels.

MOX fuels have been used in thermal reactors since the 1960s 
and today many commercial thermal reactors are loaded with one-third MOX.
Some  advanced light water designs are capable 
of accepting 100\% MOX loadings. 
In this letter we examine the antineutrino signatures from burning MOX, varying the amount and the grade of plutonium being burned.
The  potential for antineutrinos to monitor the  destruction of MOX fuel mainly lies in the fact that
fissioning $^{239}$Pu only emits about 42\% (65\%) as many antineutrinos per fission as does $^{238}$U ($^{235}$U), Table 1.
Thus, the magnitude of the emitted antineutrino flux and its variation with the fuel burnup could, in principle, verify that plutonium is the dominant actinide being burned, provided that the reactor power is known. 
Here we compare the antineutrino signals for MOX loadings of 33.3\% and 100\% of the total reactor core, either reactor-grade or weapons-grade plutonium.
In all cases, the MOX is taken to be 5.3\% PuO$_2$ and 94.7\% U$_{natural}$O$_2$. 
When the MOX fuel is 33.3\% of the core, the remaining fuel is taken to be fresh 2.56\% enriched UO$_2$. 
 
\section{Reactor Burn Simulations}
Our reactor simulations are carried out using the Monteburns code\cite{monteburns}, which  couples the Monte Carlo neutron transport code MCNP \cite{MCNP} to
the burn code CINDER`90 \cite{CINDER90}; the latter uses 63 neutron energy groups and tracks up to 3400 nuclides, including 638 isomers. 
As a benchmark, we first consider a standard  H.B. Robinson Unit 2 (HBR2) PWR fuel assembly that had 2.56\% enriched fresh LEU and no MOX fuel. 
Samples from fuel assemblies have been assayed and the spent fuel isotopics measured in detail \cite{barner}.
The sample chosen in the present work, assembly BO-5 sample N-9B-N, was irradiated to a burnup of 23.81 GWd/MTU and the detailed power history recorded.
The power history involved four burn cycles, with shutdowns of the order of 1-2 months between cycles. 
We reproduce the reported  history using variable concentrations of burnable boron poison rods over the four burn periods. We then predict spent fuel isotopic inventories for the uranium and plutonium isotopes  within about 5\% of measurement.

There are many MOX loadings that could be considered, and in the present study we consider four specific loadings. In 
the first two,  the entire reactor core is assumed to be MOX fuel, one reactor-grade (RG) and 
one weapons-grade (WG) plutonium. 
The initial isotopics are given in Table 1. The second two cases assume that
these MOX fuels make up one-third of the core; 
the remaining two-thirds of the fuel is assumed to be fresh 2.56\% enriched LEU, the same as in the HBR2 test case.
In simulating the MOX fuels,  the average power history of the HBR2 assembly is retained.  
\begin{table}
\begin{tabular}{|l|l|l|l|l|l|l|l|}\hline
Isotope (\%)&$^{235}$U&$^{238}$U&$^{238}$Pu&$^{239}$Pu&$^{240}$Pu&$^{241}$Pu&$^{242}$Pu\\\hline
Weapons&0.679&93.62&-&5.0&0.3&-&-\\
Reactor&0.679&93.62&0.203&2.653&1.373&0.535&0.535\\
$\int\sigma(E_{\overline{\nu}})S(E_{\overline{\nu}})dE_{\overline{\nu}}$ &6.61&10.10&-&5.97&-&6.1&-\\\hline 
\end{tabular}
\caption{Initial isotopics for both MOX fuels. The last row lists 
the antineutrino detection cross section per fission for each actinide in units of 10$^{-43}$ cm$^2$\cite{French, French1}, eq. 1.} 
\label{table: Iso}
\end{table}
\par\noindent{\it \underline {Entire MOX  Cores}}: 
Fig. 1 displays the nuclide fission contributions during burnup of pure RG and WG MOX fuels.
As expected,  the largest contribution to the fissions is $^{239}$Pu for all burnups considered.
The main features distinguishing the RG versus WG fuels is the 
relative importance of $^{239}$Pu versus $^{241}$Pu. In the case of burning RG plutonium, 
$^{241}$Pu accounts for $\sim15-25\%$ of the fissions, with $^{239}$Pu representing a maximum 67\%
of the fissions.
\begin{figure}
\vspace*{0.25cm}
\noindent
\includegraphics[width=3.3in]{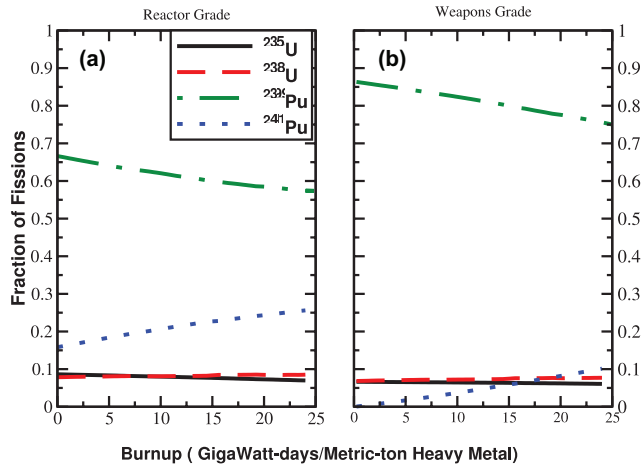}
\caption{ Fission fraction for $^{235, 238}$U, and $^{239, 241}$Pu as a function of burnup for pure  MOX fuels.   Panel (a)  is for RG and (b) for WG plutonium. The two grades of Pu are distinguished the relative importance of $^{239}$Pu and $^{241}$Pu. }
\end{figure}
\par\noindent {\it \underline {Partial MOX Core Loadings}}: 
The fission fractions for loadings involving one-third  MOX and two-thirds fresh LEU fuel are shown in Fig. 2.
 The fractions are quite distinct from those involving cores with 100\% MOX fuel.
For the partial MOX loadings, $^{235}$U and $^{239}$Pu are both dominant sources of fission.  
The two plutonium grades are  most distinguishable
by the relative importance of $^{241}$Pu, and by the value of the 
burnup at which $^{239}$Pu overtakes $^{235}$U
as the single largest contribution to the fissions. 
The fraction of fissions from $^{238}$U remains approximately constant throughout the burn and is 7-9\% of the fissions. 
\begin{figure}\vspace*{0.25cm}
\includegraphics[width=3.3in]{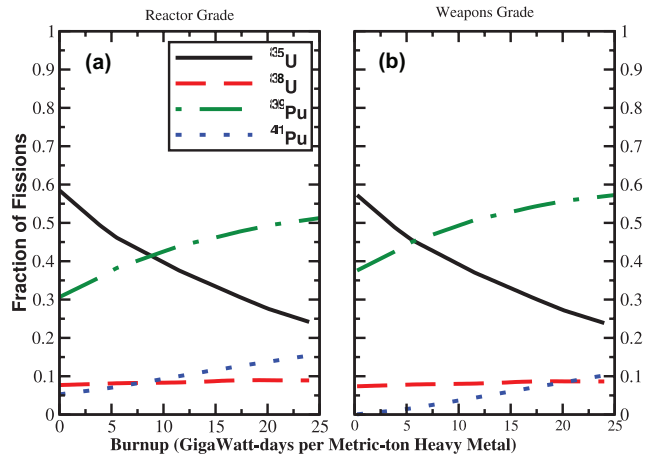}
\caption{ Fission fraction for $^{235, 238}$U, and $^{239, 241}$Pu as a function of 
burnup for 33.3\% MOX and 66.7\% LEU. Panel (a) is RG and (b) WG plutonium.}
\end{figure}
\section{Antineutrino Signals}
Approximately 5-7 antineutrinos are emitted per fission, depending on the fissioning nucleus. These antineutrinos all escape
from the reactor and they cannot be screened. 
The antineutrinos are produced in the beta-decay of the fission fragments. Their energies are in the range $\sim$0-10 MeV, and those
with energies above 1.8 MeV can be detected by the inverse beta decay reaction on the proton
($\overline{\nu}_e+p\rightarrow n +e^+$), with a cross section of the order of $5\times10^{-43} cm^2$ \cite{Borovoi1, French, French1}. 
The differences in the fission fragment distributions
for each actinide results in significant differences in the corresponding antineutrino spectra. We use the recently improved spectra of Mueller {\it et al.}\cite{French} for all isotopes.
Of course,  the overall magnitude of the emitted antineutrino flux is controlled by the power density and the size of the core,
and antineutrino safeguarding is only possible with independent knowledge  of the reactor power. 

The  differences in the antineutrino signals for the different MOX loadings, can be displayed 
by introducing an {\it effective antineutrino detection cross section per fission} for each fuel. 
\begin{equation}
\sigma_{eff}^j = \Sigma_i a_i^j\int\sigma(E_{\overline{\nu}})S_i(E_{\overline{\nu}})dE_{\overline{\nu}}\;\;,
\end{equation}  
where $a_i^j$ is the fission fraction for fuel type $j$ and each isotope $i=^{235}$U, $^{238}$U, $^{239}$Pu, $^{241}$Pu, $S_i(E)$ are the corresponding aggregate antineutrino spectra \cite{French}, and $\sigma(E)$ is the cross section for antineutrino detection on the proton.
The values of $\int\sigma(E_{\overline{\nu}})S(E_{\overline{\nu}})dE_{\overline{\nu}}$ for each actinide are listed in Table. 1.
The {\it effective} cross sections of eq. 1 vary with the burnup for each fuel type as shown in Fig. 3, and it is clear that
the MOX fuels are quite distinguishable from LEU. 
\begin{figure}\vspace*{0.25cm}
\includegraphics[width=3.2in]{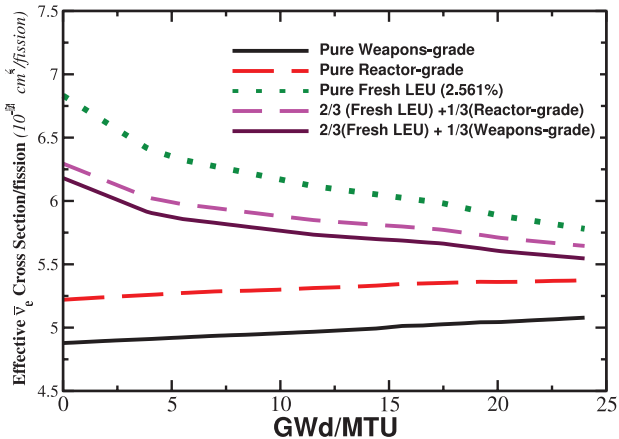}
\caption{The effective antineutrino detection cross section per fission for each fuel type. 
For 100\% MOX  the effective cross sections {\it increase} with the burn, while for 
all loading with significant LEU they {\it decrease} with the burn. 
}
\end{figure}
\begin{figure}\vspace*{0.25cm}
\includegraphics[width=3.3 in]{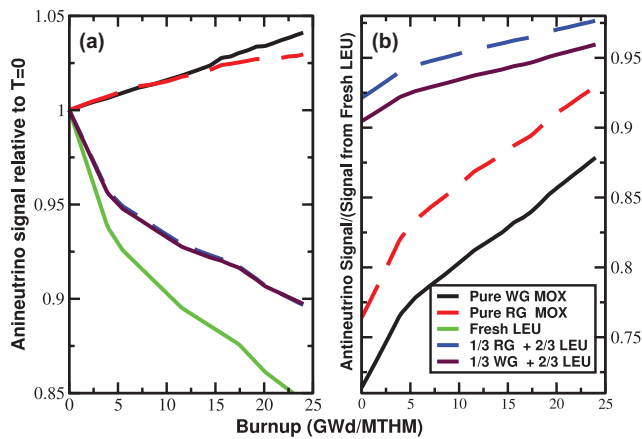}
\caption{Panel (a): The antineutrino signal relative to that at the start of the burn. Panel (b): The ratio of the antineutrino signal for MOX fuels relative to that for fresh LEU.} 
\end{figure}
The information in Fig. 3  can also be displayed in terms of the change in the expected signal 
with the fuel burnup  relative to the signal at the beginning of the burn cycle. Alternatively, if a known fuel has been monitored to calibrate the antineutrino detector, 
the ratio of the MOX signals to this known signal is useful.
These different ratios are displayed in Fig. 4, where we take the ``known'' fuel to be fresh 2.56\% LEU. 
As can be seen, pure and partial MOX fuel loadings are quite distinguishable from pure LEU. 
The larger the fraction the MOX fuel is of the reactor core, the more likely it is that the grade of plutonium can be deduced.   
\section{Summary}We investigate the antineutrino signals for four reactor core loadings of MOX; 
100\% MOX with RG or WG plutonium, and 33.3\% MOX with RG or WG plutonium and 66.6\% fresh LEU. 
In all cases we find that the antineutrino signals
are quite distinct from that expected for pure LEU fuel burning with the same power density. 
The signals are distinguishable by the combination of their magnitudes and their rate of change
with fuel burnup. In comparing the two different grades of plutonium, we find that their rate of change with respect to burnup is similar, but that the magnitude
of the signals for a given power density are different.
This difference is maximum for reactor core loadings of 100\% MOX, for which the RG signal is about 7\% larger than that for WG.  
When the MOX represents one-third of the core,
the signal difference between RG and WG is reduced to about 2\%.
If the thermal power is known, the overall uncertainty in the antineutrino flux emitted from the reactor 
is about 5\% .
This, together with the calculations presented here, suggests that burning MOX fuels containing plutonium would be quite
detectable using antineutrino monitoring. 
Distinguishing the grade of plutonium would be difficult unless the MOX  
represents the majority of  the core.

Finally, we note that all of the calculations presented here are for fresh LEU and MOX fuels. In practice, a reactor core
may be composed of fuel that is fresh, cycled through one year, two years, etc. The antineutrino signals for these
will change depending on the core composition. In all cases, the declared MOX content would be most easily verified by
taking a ratio of the antineutrino signal to that of a known fuel.
\section{ Acknowledgments}
This work was funded in part by the Los Alamos Natl. Lab. Laboratory Directed Research Development Program and
in part by  DOE NNSA's NA-22 Program.


\begin{thebibliography}{66}
\bibitem{Borovoi1} A.A. Borovoi and L.A., Mikaelyan, Soviet Atomic Energy {\bf 44}, (1978) 589; V.A. Korovlin, {\it et al.}, Soviet Atomic Energy {\bf 65}, (1988) 712; A.A. Borovoi, D.M. Vladimoirov, S.L. Gavrilov, S.L. Zvered, M.V. Lyutostanski, S. Yu, Soviet Atomic Energy {bf 70} (1991) 476; A. Bernstein, A.Wang, G. Grata, J. Applied Physics {\bf 91} (2002) 4672; M.M. Nieto, A.C. Hayes, W.B. Wilson, C.M. Teeter, W.D. Stanbro, Nucl. Sci. Eng. {\bf 149} (2005) 270; A. Bernstein, N. Bowden, A. Misner, T. Palmer, J. Applied Physics {\bf 103}, (2008) 07490; A. Porta, J. Phys. Conf. Series, {\bf 203} (2010) 01209
\bibitem{monteburns} D.I. Poston and H.R. Trellue,  LA-UR-98-2718, Los Alamos Natl. Lab. (June 1998)., and
H.R. Trellue, 
LA-1351-T, Los Alamos Natl. Lab., (Dec. 1998).

\bibitem{MCNP} J.F. Briesmeister, Editor,  LA-12625-M, Los Alamos Natl. Lab. (March 1997). 

\bibitem{CINDER90} W.B. Wilson, T.S. Cowell, T.R. England, A.C. Hayes, and P. Moller,  LAUR-07-8412, Los Alamos Natl. Lab., (Dec. 2007).

\bibitem{barner} J.O Barner,  PNL-5109 Rev.1 UC-70, Pacific Northwestern Laboratory, (June 1985).

\bibitem{French}Th. A. Mueller, D. Lhuillier, M. Fallot, A. Letourneau, S. Cormon, M. Fechner,
L. Giot, T. Lasserre, J. Martino, G. Mention, A Porta, and  F. Yermia, Phys. Rev. {\bf C 83} (2011) 054615.
\bibitem{French1} G. Mention, M. Fechner, Th. Lassere, TH. A. Mueller, D. Lhuillier, M. Cribier, and A. Letourneau, Phys. Rev. {\bf D 83}, (2011) 073006.
\end{thebibliography}
\end{document}